\documentclass[seceq,preprint]{ptptex}

\usepackage{graphicx,amsmath}
\usepackage{dcolumn}
\usepackage{bm}
\preprintnumber[5cm]{
\baselineskip0.7cm%<-- [..]: optional width of preprint # column.
YITP-08-61\\
AP-GR-61\\
OCU-PHYS-302\\
APCTP Pre2008-05
}
\title{%        %You can use \\ for explicit line-break
Solving the Inverse Problem with Inhomogeneous Universes}

\author{%       %Use \scshape  for the family name
Chul-Moon \textsc{Yoo}$^{1,3}$, Tomohiro \textsc{Kai}$^2$
and 
Ken-ichi \textsc{Nakao}$^2$
}
\inst{%         %Affiliation, neglected when [addenda] or [errata]
$^1$Yukawa Institute for Theoretical Physics, \\
Kyoto University,
Kyoto 606-8502, Japan\\
\medskip
$^{2}$Department of Physics,~Graduate School of Science, \\
~Osaka City University, Osaka 558-8585,~Japan\\
\medskip
$^{3}$Asia Pacific Center for Theoretical Physics, \\
Pohang University of Science and Technology, Pohang 790-784,~Korea\\

\recdate{%      %Editorial Office will fill in this.
July 7, 2008}

}
\abst{%         %this abstract is neglected when [addenda] or [errata]
We construct the Lema\^itre-Tolman-Bondi (LTB) dust universe 
whose distance-redshift relation is equivalent to 
that in the concordance $\Lambda$ cold dark matter ($\Lambda$CDM) 
cosmological  
model. 
In our model, the density 
distribution and velocity field are not homogeneous,  
whereas the big-bang time is uniform, which implies that the  
universe is homogeneous at its beginning. 
We also study the effects of local clumpiness in the density distribution 
as well as the effects of large-scale inhomogeneities on the 
distance-redshift relation, and 
show that these effects may reduce the amplitude of large-scale 
inhomogeneities necessary for having 
a distance-redshift relation that is the same as that of 
the concordance $\Lambda$CDM universe.  
We also study the temporal variation of the cosmological redshift 
and show that, by the observation of this quantity, 
we can distinguish our LTB universe model from 
the concordance $\Lambda$CDM model, even if their redshift-distance relations  
are equivalent to each other. 
}

\begin{document}
\maketitle

\section{Introduction}

$\Lambda$ cold dark matter ($\Lambda$CDM) models 
have achieved wide acceptance as 
concordance models due to the results of
observations over the last decade. 
In particular, cosmic microwave background (CMB)\cite{Spergel:2003cb} and 
supernovae (SNe)\cite{Riess:1998cb,Perlmutter:1998np,Knop:2003iy,Riess:2004nr} 
observations have played critical roles in this acceptance. 
The isotropy of our universe is strongly supported by 
the CMB observations. 
Thus, if we assume that our universe is homogeneous, 
results from SNe observation suggest that 
the volume expansion of our universe is accelerating. 
The accelerating expansion of a 
homogeneous and isotropic universe means 
the existence of exotic energy components, the so-called dark energy, 
within the framework of general relativity (GR). 
If we consider dark energy as 
a perfect fluid, it has negative pressure. 
One of the candidates for dark energy is the cosmological constant. 
However, there are several crucial problems regarding the 
existence of the cosmological constant or other 
dark energy candidates (see, for example, Ref.\citen{Kolb:2007gb}). 
No one knows the origin of dark energy and 
there has not yet been a conclusive illustration of dark energy. 

There are other possibilities that explain the CMB and SNe observations. 
observations. 
In the above logic, we have imposed two assumptions. 
One is validity of GR at all cosmological scales. 
The other is the homogeneity of our universe. 
Hence, in order to describe the 
observational results without dark energy, we need to 
discard GR or homogeneous models of the universe. 
In this paper, we attempt to describe the observational results 
without dark energy or any modification of gravitational theory, 
but with inhomogeneous universe models. 

The basic idea is that we are in a large underdense region, 
i.e., a large void; we reject the Copernican principle, 
which states that we live at a typical position in the universe. 
Pioneering works include those Zehavi et al.\cite{Zehavi:1998gz} 
in 1998 and Tomita\cite{2000ApJ...529...38T,Tomita:2000jj,Tomita:2001gh} 
in 2000 and 2001. 
Zehavi et al. analyze early SNe data and suggested that 
such a large void might exist around us without dark energy. 
Tomita proposed the void universe model and discussed 
the possibility of explaining the 
observed magnitude-redshift relations of SNe 
\cite{2000ApJ...529...38T,Tomita:2000jj,Tomita:2001gh}. 
There are several works in the same direction 
\cite{Celerier:1999hp,Iguchi:2001sq,Bolejko:2005fp,
Chung:2006xh,Enqvist:2006cg,Garfinkle:2006sb,
Kasai:2007fn,Biswas:2007gi}
(See also the reviews.\cite{Celerier:2007jc,Enqvist:2007vb,Mattsson:2007tj}). 
In these works,  Lema\^itre-Tolman-Bondi (LTB) 
solutions\cite{Lemaitre:1933gd,Tolman:1934za,Bondi:1947av} are often employed. 
LTB solutions are exact solutions of the Einstein equations, which 
describe the dynamics of a spherically symmetric dust fluid and  
are useful for constructing universe models with a spherically symmetric void. 

Recently, several authors have discussed the possibility of explaining the 
CMB observations\cite{Alnes:2005rw,Alexander:2007xx,GarciaBellido:2008nz}. 
They used asymptotically homogeneous LTB models and 
reported that these models may be consistent with 
the observed CMB anisotropy. 
In order to obtain more precise predictions
on whether we are located near the center of a large void, we need to study 
the evolution of nonspherical density perturbations 
in the LTB universes\cite{Zibin:2008vj}. 
Other methods of observationally investigating the inhomogeneity of our 
universe have been proposed in several papers
\cite{Clarkson:2007pz,Caldwell:2007yu,Uzan:2008qp,
Bolejko:2008cm,Clifton:2008hv}. 

The aforementioned works are 
very important for strengthening the observational  
foundation of physical cosmology. At present, the Copernican principle 
is not based on sufficient observational facts.  
However, by virtue of the recently improved observational technologies, 
we might reach the stage at which we are able to investigate through
observation whether our location in the universe is unusual or not . 
In order to test inhomogeneous universe models by observations, 
it is important 
to know the types of inhomogeneous universe models 
that can be used to explain current observations, and to reveal what 
predictions are given by these universe models. 
This is the subject of this paper.  

The inverse problem using LTB universe models is 
a useful method for investigating the possibility 
of explaining observational results with inhomogeneities in the universe 
\cite{Mustapha:1998jb,Iguchi:2001sq,Chung:2006xh,Vanderveld:2006rb}. 
LTB dust solutions contain three  arbitrary functions of 
the radial coordinate: the mass function $M$, the big-bang time 
$t_{\rm B}$ and the curvature function $k$,  
and the inverse problem means determining 
$M$, $t_{\rm B}$ and $k$ so that 
a given distance-redshift relation is realized on the past light-cone 
of an observer. 
In order to specify the three arbitrary functions, we need three conditions. 
One of these conditions corresponds to the choice of the radial coordinate and thus 
has no physical meaning.  Hence, one more condition in addition to 
the distance-redshift relation is necessary. 

Iguchi et al.\cite{Iguchi:2001sq} showed that 
the distance-redshift relation in the standard $\Lambda$CDM model can be 
reproduced in LTB dust universe models. 
When they solved the inverse problem, they imposed two kinds of the 
additional condition:
one is a uniform big-bang time $t_{\rm B}=0$ 
and the other is a vanishing curvature function $k=0$. 
It should be noted that their models failed to reproduce the distance-redshift 
relation for $z\gtrsim 1.7$. 
The reason why they could not continue their computation beyond $z\sim1.7$ 
is the existence of the ``critical point'' 
discussed by Vanderveld et.al\cite{Vanderveld:2006rb}. 
They reported the difficulty in solving the inverse problem and 
concluded that it is unlikely that a solution to the 
inverse problem can be found in all the redshift domain. 
However, Tanimoto and Nambu 
recently solved the inverse problem in all the redshift 
domain with a nonuniform $t_{\rm B}$\cite{TNpriv}. 
In this paper, we show that 
it is possible to obtain a solution to the inverse problem with 
the condition of the uniform big-bang time $t_{\rm B}$. 
\footnote{
Mustapha et al.\cite{Mustapha:1998jb} proposed the 
following theorem
:{\it Subject to the conditions of Appendix B in Ref.\citen{Mustapha:1998jb} 
for any given isotropic observation of apparent luminosity $l(z)$ and 
number count $n(z)$ with any given source evolution function $\hat L(z)$ 
and total density over source number density $\hat m(z)$, 
a set of LTB functions can be found to make the LTB observational 
relations fit the observations. } 
It is nontrivial whether the same statement holds for any given set of some
quantities different from $l(z)$, $n(z)$, $\hat{L}(z)$ and $\hat{m}(z)$.
In this paper, the big-bang time $t_{\rm B}(r)$ is given as one of
the quantities.
Thus, the situation is different from the case in Ref.\citen{Mustapha:1998jb}. }
In terms of perturbation theory, the inhomogeneity of the big-bang time corresponds to 
the decaying mode. 
The condition of uniform big-bang time might guarantee 
the consistency of the present model with the inflationary scenario, 
since a universe that experiences inflation 
is almost homogeneous immediately after the inflationary period is over. 

In this paper, we also discuss the effects of local clumpiness. 
In addition to the effect of the large-scale void structure, 
small-scale clumpiness 
may affect the observed distance-redshift relation. 
If there are clumpy objects such as galaxies in the foreground of 
light sources, data for the apparent luminosities of these sources 
are often contaminated with gravitational lensing or absorption. 
For this reason, supernova teams specifically search for 
objects with a minimum amount of intervening 
material in the foreground\cite{Pain:2002wj,Mattsson:2007tj}. 
The observed sources might be biased by this selection, since 
the light of the observed sources might propagate in a lower-density region 
of our universe. 
The simplest way to take account of this effect is to 
introduce the so-called smoothness parameter 
$\alpha$, which is the ratio 
of the smoothly distributed matter density except for clumps to the 
mean energy density $\rho$ for all matter, 
assuming that the energy density on the paths of observed light rays 
is given by $\alpha\rho$
\cite{1969ApJ...155...89K,1973ApJ...180L..31D,Kantowski1998}. 
Tomita has studied the effects of small-scale clumpiness in a local void
model by introducing the smoothness parameter $\alpha$. 
In this paper, we also investigate the corresponding 
effects by introducing
$\alpha$.

The time derivative of the cosmological redshift 
is a very important observable 
quantity in distinguishing LTB models from 
concordance $\Lambda$CDM universe models. 
We study it and show that it gives a 
criterion for observationally deciding whether the universe is described by 
an LTB model with uniform big-bang time. 

This paper is organized as follows. 
In \S\ref{sec2}, we briefly review LTB dust universes 
and list basic equations that we have to solve. 
The temporal variation of the cosmological redshift of 
comoving sources is discussed in \S\ref{sec3}. 
Then, we show details of our numerical method, 
particularly, focusing on singular points of the equations at 
the center and the critical point in \S\ref{sec4}. 
Numerical results are given in \S\ref{sec5}. 
Section\ref{sec6} is devoted to a summary and discussion. 

Throughout this paper, we use the unit of $c=G=1$, 
where $c$ and $G$ are the speed of light and the 
gravitational constant, respectively. 
%%%%%%%%%%%%%%%%%%%%%%%%%%%%%%%%%%%%%%%%%%%%%%%%%%%%%%%%%%%%%%%%
\section{Basic equations}
\label{sec2}
%%%%%%%%%%%%%%%%%%%%%%%%%%%%%%%%%%%%%%%%%%%%%%%%%%%%%%%%%%%%%%%%
\subsection{LTB dust universe}
%%%%%%%%%%%%%%%%%%%%%%%%%%%%%%

LTB solutions are exact solutions to the Einstein equations, which 
describe the dynamics of a spherically symmetric 
dust fluid and whose line element is written in the form 
\begin{equation}
ds^2=-dt^2+\frac{\left(\partial_r R(t,r)\right)^2}{1-k(r)r^2}dr^2
+R^2(t,r)d\Omega^2, \label{eq:metric}
\end{equation}
where $k(r)$ is an arbitrary function of the radial coordinate $r$. 
The LTB solutions include 
homogeneous and isotropic universes as special cases;  
in this case, $k$ is a constant called the curvature parameter 
in the appropriate gauge, and thus  
we call it the curvature function. 
The stress-energy tensor of the dust is given by 
\begin{equation}
T^{\mu\nu}
=\rho(t,r) u^\nu u^\nu, \label{eq:enmomedust}
\end{equation}
where $\rho(t,r)$ is the rest mass density of the dust and 
$u^\mu=\delta^\mu_0$ 
is the 4-velocity of a dust particle. 
The Einstein equations lead to the equations for the areal radius 
$R(t,r)$ and the rest mass density $\rho(t,r)$, 
\begin{equation}
\left(\partial_t R\right)^2=-k(r)r^2+\frac{2M(r)}{R}\label{eq:fory}
\end{equation}
and
\begin{equation}
4\pi\rho=\frac{\partial_r M(r)}{R^2\partial_r R }, \label{eq:rho}
\end{equation}
where $M(r)$ is an arbitrary function of the radial coordinate $r$. 
We assume that $\rho$ is nonnegative and that 
$R$ is monotonic with respect to $r$, 
i.e., $\partial_r R>0$.

Following Tanimoto and Nambu \cite{Tanimoto:2007dq}, 
the solution to Eqs. (\ref{eq:fory}) and (\ref{eq:rho}), 
which represents the expanding universe, is written in the form  
\begin{eqnarray}
R(t,r)&=&(6M(r))^{1/3}(t-t_{\rm B}(r))^{2/3}\mathcal S(x), 
\label{eq:YS}\\
x&=&k(r)r^2\left(\frac{t-t_{\rm B}(r)}{6M(r)}\right)^{2/3}, \label{eq:defx}
\end{eqnarray}
where $t_{\rm B}(r)$ is an arbitrary function of the radial coordinate $r$. 
The function $\mathcal S(x)$ is defined as
\begin{equation*}
\mathcal S(x)=\frac{\cosh\sqrt{-\eta}-1}{6^{1/3}(\sinh\sqrt{-\eta}
-\sqrt{-\eta})^{2/3}}~~,~~x=\frac{-(\sinh\sqrt{-\eta}-\sqrt{-\eta})^{2/3}}{6^{2/3}}
~~{\rm for}~~x<0.
\end{equation*}
\begin{equation}
\mathcal S(x)=\frac{1-\cos\sqrt{\eta}}{6^{1/3}(\sqrt{\eta}
-\sin\sqrt{\eta})^{2/3}}~~,
~~x=\frac{(\sqrt{\eta}-\sin\sqrt{\eta})^{2/3}}{6^{2/3}}~~{\rm for}~~x>0.
\label{eq:defS}
\end{equation}
\begin{equation*}
\mathcal S(0)=\left(\frac{3}{4}\right)^{1/3}. 
\end{equation*}
$\mathcal S(x)$ is analytic in the domain 
$x<x_c\equiv (\pi/3)^{2/3}$. 
Some characteristics of the function $\mathcal S(x)$ 
are given in Appendix \ref{sec:funcS} and Ref.\citen{Tanimoto:2007dq}. 
We can easily see from the above 
equations that the areal radius  $R$ vanishes at $t=t_{\rm B}(r)$. 
Thus, 
the function $t_{\rm B}(r)$ is called the big-bang time. 

Following Refs.\citen{Enqvist:2006cg} and \citen{Enqvist:2007vb}, 
we define the local Hubble function by 
\begin{equation}
H(t,r)=\frac{\partial_t R}{R}
\end{equation}
and the density-parameter function of dust as 
\begin{equation}
\Omega_{\rm M}(t,r)=\frac{2M(r)}{H(t,r)^2R(t,r)^3}. 
\end{equation}
In LTB universes, we can define another expansion rate of the spatial length 
scale, the so-called longitudinal expansion rate, by 
\begin{equation}
H^L(t,r)=\frac{\partial_t\partial_rR}{\partial_rR}. 
\label{eq:deflongh}
\end{equation}
In the case of homogeneous and isotropic universes, 
$H^L$ agrees with $H$. 
Thus, if we can measure the difference between $H$ and $H^L$, 
it can be used as 
an indicator of the inhomogeneity 
in the universe\cite{Clarkson:2007pz}.

\subsection{Conditions and equations to determine arbitrary functions}

As shown in the preceding subsection, the LTB solutions 
have three arbitrary functions: $k(r)$, $M(r)$ and $t_{\rm B}(r)$. 
We have one degree of freedom to rescale the radial coordinate $r$. 
To fix one of the three functional degrees of freedom corresponds to 
fixing the gauge freedom 
of this rescaling, and it will be shown later how to fix it. 
The remaining two functional degrees of freedom are fixed 
by imposing the following physical conditions. 

\begin{itemize}
\item{Uniform big-bang time $t_{\rm B}=0$.}
\item{The angular diameter distance $D(z)$ 
is equivalent to that in the $\Lambda$CDM universe in 
all the redshift domain except in the vicinity 
of the symmetry center 
in which $D$ is appropriately set so that the regularity 
of the spacetime geometry is guaranteed. 
}
\end{itemize}
Here we stress that it is our primary purpose to find an LTB model that
fits the observed distance-redshift relation well. 
Thus, it does not matter
that the distance-redshift relation does not agree with that of $\Lambda$CDM
model only in the vicinity of the symmetry center.

In order to determine $k(r)$ and $M(r)$ from the above conditions, we 
consider a past-directed outgoing radial null geodesic 
that emanates from the observer at the center. 
This null geodesic is expressed in the form
\begin{eqnarray}
t&=&t(\lambda), \\
r&=&r(\lambda), 
\end{eqnarray}
where $\lambda$ is an affine parameter. 

We assume that the observer is always located at the symmetry center 
$r=0$ and observes the light ray at $t=t_0$. 
In order to fix the gauge freedom to rescale the radial coordinate $r$, 
we adopt the light-cone gauge condition that the relation  
\begin{equation}
t=t_0-r 
\label{eq:gaucon}
\end{equation}
is satisfied along the observed light ray. 

Then the basic equations to determine $k$ and $M$ are given as follows:

\begin{enumerate}

\item{\bf Null condition}

By virtue of the light-cone gauge condition, the null condition 
on the observed light ray takes the very simple form of  
\begin{equation}
\partial_r R=\sqrt{1-kr^2}. 
\label{eq:nullcon}
\end{equation}

\item{\bf Definition of redshift}

The redshift is defined by
\begin{equation}
1+z=\frac{\left.u^\mu p_\mu\right|_{\rm source}}{\left.u^\mu 
p_\mu\right|_{\rm observer}}\propto p_0=-p^0=-\dot t, 
\end{equation}
where $p^\mu$ is the tangent vector of the null geodesic that corresponds 
to the observed light ray and the dot 
represents differentiation with respect to the affine parameter $\lambda$. 
By using the freedom to multiply the affine parameter by a 
constant, we can write, without loss of generality,  
\begin{equation}
\dot t=-\dot r=-\frac{1+z}{H_0}, 
\label{eq:defz}
\end{equation}
where 
\begin{equation}
H_0:=H(t_0,0), 
\end{equation}
and we have used the gauge condition (\ref{eq:gaucon}) in the first equality. 
In this normalization, the affine parameter is dimensionless.

\item{\bf Geodesic equation}

One of the geodesic equations for the radial null geodesic is given by 
\begin{equation}
(\partial_r R)\ddot t+(\partial_t\partial_r R) \dot t^2=0, 
\label{eq:nullgeo}
\end{equation}
where we have used the null condition (\ref{eq:nullcon}). 

\item{\bf Dyer-Roeder equation for the angular diameter distance}

As mentioned, in order to fix the remaining functional freedom, we 
assume the angular diameter distance $D(z)$. 
Then the Dyer-Roeder equation 
\begin{equation}
\ddot z\frac{dD}{dz}+\dot z^2\frac{d^2D}{dz^2}=
-4\pi\frac{(1+z)^2}{H_0^2}\alpha\rho D 
\label{eq:dreq}
\end{equation}
gives us one of the equations to determine the arbitrary functions 
of LTB universe models, where $\alpha$ is the smoothness 
parameter mentioned in \S1. 
The derivation for this equation is given in Appendix \ref{sec:DRdis}. 
$\alpha$ is the mass ratio of the smoothly distributed 
components of matter to all the components, and it may vary 
with time due to the formation of structures in the real universe. Therefore, 
in this paper, we assume that $\alpha$ is an input function of 
the cosmological redshift $z$.  

\end{enumerate}

Equations (\ref{eq:nullcon}), (\ref{eq:defz}), 
(\ref{eq:nullgeo}) and (\ref{eq:dreq})
are rewritten in the form of five coupled ordinary differential equations:
\begin{eqnarray}
\dot m&=&F_m(m,k,r,z,\zeta), \label{eq:dm}\\
\dot k&=&F_k(m,k,r,z,\zeta), \label{eq:dk}\\
\dot r &=&\frac{1+z}{H_0}, \label{eq:dchi}\\
\dot z&=&\frac{\zeta}{dD/dz}, \label{eq:dz}\\
\dot \zeta&=&\frac{-4\pi(1+z)^2\alpha\rho D}{H_0^2}, \label{eq:dzeta}
\end{eqnarray}
where $m$ and $\zeta$ are defined by 
\begin{equation}
m(r):=\frac{6M(r)}{r^3}
\end{equation}
and 
\begin{equation}
\zeta:=\dot z\frac{dD}{dz}, \label{eq:defzeta}
\end{equation}
respectively. From Eq. (\ref{eq:rho}), we have 
\begin{equation}
\rho
=\frac{r^2\left[3(1+z)m+H_0rF_m(m,k,r,z,\zeta)\right]}{24\pi (1+z)R^2\sqrt{1-kr^2}}. 
\end{equation}
The derivation of these equations is shown in Appendix \ref{sec:cal}. 

\section{Temporal variation of the cosmological redshift}
\label{sec3}

The temporal variation of the cosmological redshift 
will give us crucial information about the acceleration of cosmic volume 
expansion or inhomogeneities in our universe\cite{Lake}. 

\subsection{Homogeneous and isotropic universe}

The line element of homogeneous and isotropic universes is given by
\begin{equation}
ds^2=-dt^2+a^2(t)\left(d\chi^2+\Sigma(\chi)d\Omega^2\right),
\end{equation}
where $\Sigma(\chi)=\sin\chi$ for a closed universe, $\Sigma(\chi)=\chi$ 
for a flat universe and $\Sigma(\chi)=\sinh\chi$ for a open universe. 
In this subsection, we assume that the universe is 
filled with dust and dark energy 
characterized by the linear equation of state $p=w\rho$, 
where $p$ is the pressure, $\rho$ is the energy density and $w$ is a constant 
less than $-1/3$.  We assume that the energy densities of both dust and dark 
energy are nonnegative. 

The Einstein equations lead to 
\begin{eqnarray}
\left(\frac{1}{a}\frac{da}{dt}\right)^2&=&\Omega_{\rm M0}
\left(\frac{a_0}{a}\right)^3
+\Omega_{\rm X0}\left(\frac{a_0}{a}\right)^{3(1+w)}+(1-\Omega_{\rm M0}-\Omega_{\rm X0})
\left(\frac{a_0}{a}\right)^2, \label{eq:F-eq}
\end{eqnarray}
where $a_0$ is the scale factor at $t=t_0$, and $\Omega_{\rm M0}$ and 
$\Omega_{\rm X0}$ are the density parameters of the dust and the 
dark energy, respectively. Note that, by assumption, both 
$\Omega_{\rm M0}$ and $\Omega_{\rm X0}$ are nonnegative.  
As is well known, the cosmological redshift $z$ 
of a light signal emitted from a comoving source is given by
\begin{equation}
z=\frac{a(t_0)}{a(t_{\rm e})}-1,
\end{equation}
where $t_{\rm e}$ is the time when the light is emitted from the source. 
The temporal variation of $z$ of a comoving source is then given by
\begin{equation}
\Delta z
=\frac{a(t_0+\Delta t_0)}{a(t_{\rm e}+\Delta t_{\rm e})}
-\frac{a(t_0)}{a(t_{\rm e})}
\sim H_0\left(1+z-\frac{H_{\rm e}}{H_0}\right)\Delta t_0,
\end{equation}
where $\Delta t_{\rm e}=\Delta t_0/(1+z)$ and 
$H_{\rm e}$ is the Hubble parameter when the light is emitted from the source. 
Substituting Eq. (\ref{eq:F-eq}) into the above equation, we have
\begin{equation}
\frac{dz}{dt_0}=H_0(1+z)\left[1
-\sqrt{1+\Omega_{\rm M0}z+\Omega_{\rm X0}\left\{(1+z)^{1+3w}-1\right\}}\right]. 
\label{eq:dzdt-CDM}
\end{equation}
Note that, $\Omega_{\rm M0}z$ is nonnegative, whereas 
$\Omega_{\rm X0}\{(1+z)^{1+3w}-1\}$ is nonpositive due 
to the assumption $w<-1/3$. 
Thus, if the dust is a dominant component of the universe, i.e., 
$\Omega_{\rm M0} \gg \Omega_{\rm X0}$, then $dz/dt_0$ is negative. 
By contrast, in the 
case of a universe dominated by dark energy, i.e., 
$\Omega_{\rm X0} \gg \Omega_{\rm M0}$, then  
$dz/dt_0$ is positive. 
Therefore, the measurement of the temporal variation of the
cosmological redshift $z$ will give us 
crucial knowledge about the equation of state if the universe is 
homogeneous and isotropic. 

\subsection{LTB universe}

In order to obtain the time derivative of the cosmological 
redshift $z$ in LTB universe models, we consider another past-directed 
outgoing radial null geodesic that is infinitesimally close to the 
null geodesic considered in the preceding section,
\begin{equation}
t=t_{\rm b}(\lambda)+\delta t(\lambda)~~~~~{\rm and}~~~~~
r=r_{\rm b}(\lambda)+\delta r(\lambda), \label{eq:1st}
\end{equation}
where $t_{\rm b}(\lambda)$ and $r_{\rm b}(\lambda)$ denote the 
null geodesic with the initial condition $r(0)=0$ at $t(0)=t_0$, 
which was considered in the preceding section. 
We set the affine parameter so that the cosmological redshift is given by
\begin{equation}
\dot{t}=\dot{t}_{\rm b}+\delta\dot{t}=-\frac{1+z}{H_0}=-\frac{1+z_{\rm b}+\delta z}{H_0},
\label{redshift}
\end{equation}
where the subscript ``b'' denotes the value evaluated on the null geodesic 
$(t,r)=(t_{\rm b},r_{\rm b})$. 
Thus, we have 
\begin{equation}
\delta z=-H_0\delta\dot{t}. 
\end{equation}

Substituting Eq. (\ref{eq:1st}) into Eq. (\ref{eq:nullgeo}), 
and taking the first order of $\delta t$ and $\delta r$, we have
\begin{eqnarray}
\delta\ddot{t}+2H^L_{\rm b}\dot{t}_{\rm b}\delta\dot{t}
+\left[\frac{\partial_t^2\partial_rR}{\partial_rR}
-(H^L)^2\right]_{\rm b}\dot t_{\rm b}^2\delta t
+\left[\frac{\partial_t\partial_r^2R}{\partial_rR}-H^L
\frac{\partial_r^2R}{\partial_rR}\right]_{\rm b}\dot t_{\rm b}^2
\delta r=0.\label{eq:deltat}
\end{eqnarray}
Another equation is given by the null condition,
\begin{eqnarray}
\delta\dot{r}=-\delta\dot{t}+H^L_{\rm b}\dot t_{\rm b}\delta t+
\left[\frac{\partial_rkr^2+2kr}{2(\partial_rR)^2}
+\frac{\partial_r^2R}{\partial_rR}\right]_{\rm b}\dot t_{\rm b}\delta r. 
\label{eq:deltar}
\end{eqnarray}
The initial conditions for these equations are given by 
\begin{align}
\delta \dot t(0)&=0, \\
\delta r(0)&=0, \\
\delta t(0)&={\rm const.}.
\end{align}

The temporal variation of the cosmological redshift of a comoving source 
is given by 
\begin{equation}
\Delta z(\lambda)
=z(\lambda+\Delta\lambda)-z_{\rm b}(\lambda), \label{eq:del-z}
\end{equation}
where the infinitesimal quantity $\Delta\lambda$ satisfies the equality
\begin{equation}
r(\lambda+\Delta\lambda)=r_{\rm b}(\lambda). \label{eq:del-lambda}
\end{equation}
From Eqs. (\ref{eq:1st}), (\ref{redshift}), 
(\ref{eq:del-z}) and (\ref{eq:del-lambda}), we have, 
up to the first order of $\Delta\lambda$, 
\begin{equation}
\Delta z(\lambda)\sim\delta z(\lambda)+\dot{z}_{\rm b}(\lambda)\Delta\lambda
\sim\delta z(\lambda)
-\frac{\dot{z}_{\rm b}(\lambda)}{\dot{r}_{\rm b}(\lambda)}\delta r(\lambda).
\end{equation}
Thus, we have the derivative of the cosmological redshift $z$ with respect to 
the time of the observer as a function of the affine parameter, 
\begin{equation}
\frac{dz}{dt_0}(\lambda):=
\frac{\delta z(\lambda)}{\delta t(0)}-\frac{\dot{z}_{\rm b}(\lambda)}{\dot{r}_{\rm b}(\lambda)}
\frac{\delta r(\lambda)}{\delta t(0)}.
\end{equation}
Since we also obtain $z_{\rm b}(\lambda)$ simultaneously, we have 
the relation between $z$ and $dz/dt_0$. 
In order to solve the differential equations (\ref{eq:deltat}) and 
(\ref{eq:deltar}), 
we need to know $(\partial_r^2R)_{\rm b}$, 
$(\partial_t^2\partial_rR)_{\rm b}$ and 
$(\partial_t\partial_r^2R)_{\rm b}$, which 
are given in Appendix \ref{sec:delrs}. 
Using Eq.(\ref{eq:D2}), Eq.(\ref{eq:deltar}) can be 
solved as 
\begin{align}
\delta r+\delta t=\frac{\delta t(0)}{1+z}, 
\end{align}
where we have used Eqs.(\ref{eq:deflongh}) and (\ref{eq:nullgeo}). 

\section{Singularity at the center and critical point}
\label{sec4}
\subsection{Resolving singularity at the center}
\label{sec:4a}

Suppose that the reference angular diameter distance 
$D(z)$ is exactly the same as that of the concordance $\Lambda$CDM model with 
$(\Omega_{\rm M0},\Omega_{\Lambda0})=(0.3, 0.7)$, where $\Omega_{\Lambda0}$ is the 
density parameter of the cosmological constant. 
Then, as shown in Appendix \ref{sec:iniv}, once 
$\Omega_{\rm M}(t_0,0)$ is given, the set of solutions for the differential 
equations (\ref{eq:dm})-(\ref{eq:dzeta}) 
is uniquely determined. 
Vanderveld et al.\cite{Vanderveld:2006rb} reported that 
many of the inhomogeneous 
models that mimic observations of an accelerating universe 
contain a weak singularity at the symmetry center. 
This singularity is too weak to make the spacetime geodesically
incomplete, and thus we may accept these models as being effective. 

On the other hand, the accuracy of observations in the low-redshift
domain is not sufficient to uniquely determine the distance-redshift
relation. 
Thus, we need to assume the redshift
dependence of the angular diameter distance in the low-redshift domain.
If the $C^\infty$ model is preferred, we may assume the following input angular
diameter distance $D(z)$,
\begin{equation}
D=D_{(0.3,0.7)}(z)\left[1-\exp\left(-z^2/\delta^2\right)\right]
+D_{(\Omega_{\rm m0},0)}(z)\exp\left(-z^2/\delta^2\right), 
\label{eq:ref-D}
\end{equation}
where $\delta$ is a positive constant, whereas  
$D_{(0.3,0.7)}$ and $D_{(\Omega_{\rm m0},0)}$ are 
the angular diameter distances in the isotropic and homogeneous universe 
with $(\Omega_{\rm M0},\Omega_{\Lambda0})=(0.3, 0.7)$ and 
$(\Omega_{\rm M0},\Omega_{\Lambda0})=(\Omega_{\rm m0}, 0)$, respectively. 
$D(z)$ is almost the same as the angular diameter distance of the $\Lambda$CDM
model for $z>\delta$,
whereas, 
in the vicinity of the symmetry center, 
$D(z)$ is almost the same as that of 
the homogeneous and isotropic universe 
with 
$(\Omega_{\rm M0},\Omega_{\Lambda0})=(\Omega_{\rm m0}, 0)$. 
It should be noted that $\Omega_{\rm m0}$ is equal to $\Omega_{\rm M}(t_0,0)$ 
and can be regarded as a parameter to specify the LTB universe model. 

Also note that if our vicinity is described
well by the very smooth LTB model, the distance-redshift relation does not
agree with that of the concordance $\Lambda$CDM model in the
low-redshift domain. 
Nambu and Tanimoto have
shown that the Maclaurin series of $D(z)$ for the LTB model with a regular
symmetry center agrees with the homogeneous and isotropic 
dust-filled universe up to $z^2$. 
The angular diameter distance
(4.1) is one of the simplest assumptions that ensures the regularity at the
symmetry center and agrees with that of the concordance $\Lambda$CDM model
in the high-redshift domain.

\subsection{Critical point}

The differential equation (\ref{eq:dz}) has a singular point at which 
$dD/dz$ vanishes. 
Following Ref.\citen{Vanderveld:2006rb}, we call this point the critical point 
and denote the cosmological redshift at the critical point by $z_{\rm cr}$. 
We can require that the value of $\zeta$ should vanish at the critical point so that 
the solution is regular at this point. 
This gives a constraint on the free parameter $\Omega_{\rm m0}$.

The symmetry center $r=0$ is another regular singular point of the 
differential equations (\ref{eq:dm})-(\ref{eq:dzeta}). 
The Runge-Kutta method, which we have used, is generally unstable 
for solving ordinary differential equations toward a regular singular point 
from a regular point, and thus we start the numerical integration from 
these regular singular points; we numerically integrate 
the equations from the symmetry center $r=0$ but not to the 
critical point $z=z_{\rm cr}$, 
and we also integrate them from the critical point $z=z_{\rm cr}$ 
but not to the symmetry center $r=0$. 
Instead, specifying $\Omega_{\rm m0}$ and 
the values of $m$, $k$ and $r$ at the critical point $z=z_{\rm cr}$, 
we numerically integrate the differential equations outward 
from the symmetry center $r=0$ 
and inward from the critical point $z=z_{\rm cr}$ 
to the matching point $z=z_{\rm m}$ 
located in the domain between these singular points. 

If we fail to choose appropriate values of 
$\Omega_{\rm m0}$ and values of $m$, $k$ and $r$ 
at the critical point $z=z_{\rm cr}$, the 
resultant solutions are discontinuous at $z=z_{\rm m}$. 
Thus, we have to search for the appropriate initial values 
for $\Omega_{\rm m0}$, and $m$, $k$ and $r$ at the critical point $z=z_{\rm cr}$ 
so that the following matching conditions are satisfied: 
\begin{equation}
\begin{array}{rcl}
m|_{z=z_{\rm m}+0}&=&m|_{z=z_{\rm m}-0}, \\
k|_{z=z_{\rm m}+0}&=&k|_{z=z_{\rm m}-0}, \\
r|_{z=z_{\rm m}+0}&=&r|_{z=z_{\rm m}-0}, \\
\zeta|_{z=z_{\rm m}+0}&=&\zeta|_{z=z_{\rm m}-0}.
\end{array}
\label{eq:match}
\end{equation}
Note that if the above conditions hold, the smoothness of the solutions 
is also guaranteed, since the 
equations for these functions are first-order differential equations. 

We have searched for appropriate initial conditions 
that guarantee the matching conditions (\ref{eq:match}) 
by using the four-dimensional Newton-Raphson method. 
Using this procedure, we can uniquely obtain the solution 
if the value of $\delta$ in Eq. (\ref{eq:ref-D}) is fixed.

\section{Numerical results}
\label{sec5}

We have solved Eqs. (\ref{eq:dm})-(\ref{eq:dzeta}) by using the numerical 
procedure described in the previous section. 
In the following subsections, we express $k(r(z)$, $m(r(z))$, 
$\Omega_{\rm M}(t_0,r(z))$, $H(t_0,r(z))$, $H^L(t_0,r(z))$ and $\rho(t_0,r(z))$ as 
functions of the cosmological redshift $z$. 
We also express the time derivative of the cosmological 
redshift $dz/dt_0$ as a function of the cosmological redshift $z$ itself 
together with that of the concordance $\Lambda$CDM universe
with $(\Omega_{\rm M0},\Omega_{\Lambda0})=(0.3, 0.7)$.

\subsection{Results without local clumpiness $(\alpha=1)$}

\begin{figure}[htbp]
\begin{center}
\includegraphics[scale=0.7]{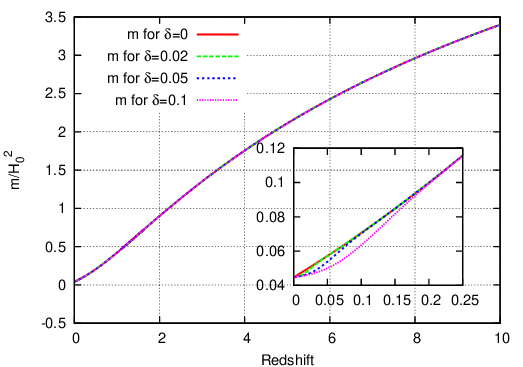}
\caption{$m(r(z))$ depicted as 
functions of the cosmological redshift for various values of $\delta$. 
}
\label{fig:m}
\end{center}
\end{figure}

\begin{figure}[htbp]
\begin{center}
\includegraphics[scale=0.7]{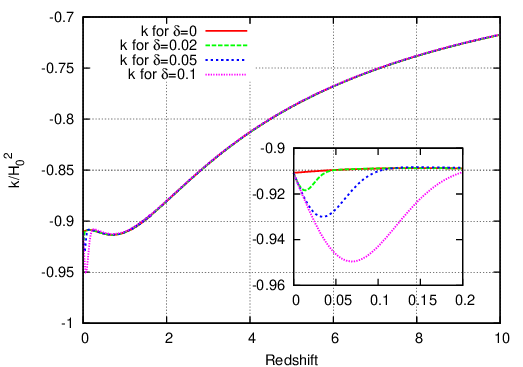}
\caption{$k(r(z))$ 
depicted as functions of the cosmological redshift for 
various values of $\delta$. 
}
\label{fig:k}
\end{center}
\end{figure}

In this subsection, we assume $\alpha=1$. 
As mentioned above, 
the solution is uniquely given by our numerical procedure 
if the value of $\delta$ in Eq. (\ref{eq:ref-D}) is fixed. 
First, we show 
$m(r(z))$ and $k(r(z))$ as functions of $z$ 
in Figs. \ref{fig:m} and \ref{fig:k}, respectively. 
The results do not strongly depend on the value of $\delta$ except when 
$z\lesssim \delta$. 
%%%%%%%%%%%%%%%%%%%%%%%%%%%<<start figure>>%%%%%%%%%%%%%%%%%%%%%%%%%%
%%%%%%%%%%%%%%%%%%%%%%%%%%%%<<end figure>>%%%%%%%%%%%%%%%%%%%%%%%%%%%
%%%%%%%%%%%%%%%%%%%%%%%%%%%<<start figure>>%%%%%%%%%%%%%%%%%%%%%%%%%%
%%%%%%%%%%%%%%%%%%%%%%%%%%%%<<end figure>>%%%%%%%%%%%%%%%%%%%%%%%%%%%

\begin{figure}[htbp]
\begin{center}
\includegraphics[scale=0.7]{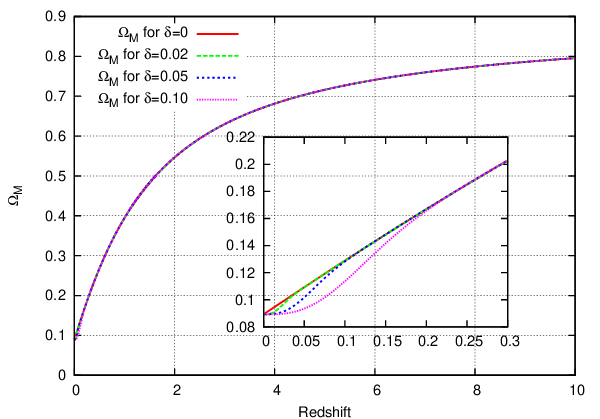}
\caption{Density-parameter functions $\Omega_{\rm M}(t_0, r(z))$ depicted as 
functions of the cosmological redshift for various values of $\delta$. 
}
\label{fig:0plotom}
\end{center}
\end{figure}

\begin{figure}[htbp]
\begin{center}
\includegraphics[scale=0.7]{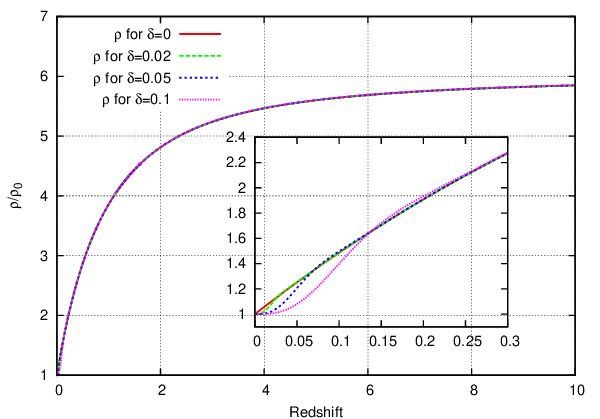}
\caption{Rest-mass densities $\rho(t_0,r(z))$ depicted as 
functions of the cosmological redshift for various values of $\delta$. 
}
\label{fig:0plotrho}
\end{center}
\end{figure}

\begin{figure}[htbp]
\begin{center}
\includegraphics[scale=0.7]{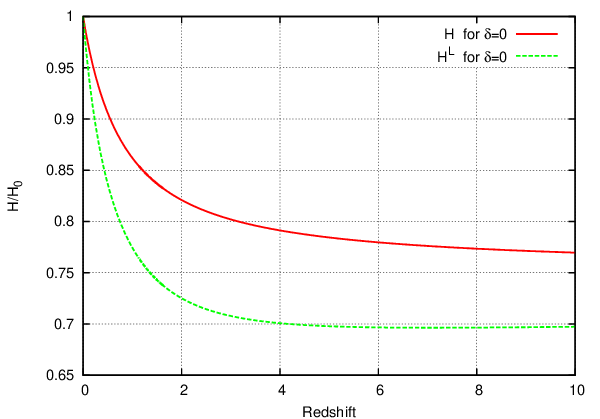}
\caption{Local Hubble function $H(t_0,r(z))$ and 
longitudinal expansion rate $H^L(t_0,r(z))$ 
depicted as functions of the cosmological redshift for $\delta=0$. 
}
\label{fig:0ploth}
\end{center}
\end{figure}

In order to determine the physical properties of the solution, 
we depict $\Omega_{\rm M}(t_0,r(z))$, $\rho(t_0,r(z))$, and 
$H(t_0,r(z))$ and $H^L(t_0,r(z))$, respectively, 
as functions of $z$ in
Figs. \ref{fig:0plotom}-\ref{fig:0ploth}. 
We can see from these figures that the resultant inhomogeneity 
is a large-scale void structure. 
%%%%%%%%%%%%%%%%%%%%%%%%%%%<<start figure>>%%%%%%%%%%%%%%%%%%%%%%%%%%
%%%%%%%%%%%%%%%%%%%%%%%%%%%%<<end figure>>%%%%%%%%%%%%%%%%%%%%%%%%%%%
%%%%%%%%%%%%%%%%%%%%%%%%%%%<<start figure>>%%%%%%%%%%%%%%%%%%%%%%%%%%
%%%%%%%%%%%%%%%%%%%%%%%%%%%%<<end figure>>%%%%%%%%%%%%%%%%%%%%%%%%%%%
%%%%%%%%%%%%%%%%%%%%%%%%%%%<<start figure>>%%%%%%%%%%%%%%%%%%%%%%%%%%
%%%%%%%%%%%%%%%%%%%%%%%%%%%%<<end figure>>%%%%%%%%%%%%%%%%%%%%%%%%%%%
We note that $H(t_0,r(z))=H^L(t_0,r(z))$ in homogeneous 
and isotropic universes, whereas 
$H(t_0,r(z))$ and $H^L(t_0,r(z))$ 
are different from each other by about 10\% 
for $2\lesssim z<10$ in the inhomogeneous case depicted 
in Fig. \ref{fig:0ploth}. 
This result means that, in order to fit the distance-redshift relation of
the LTB model with  observations that almost agree with that predicted by
the concordance $\Lambda$CDM universe with 
$(\Omega_{\rm M0},\Omega_{\Lambda0})=(0.3, 0.7)$, the scale of the
inhomogeneity should be at least a few Gpc.

\subsection{Results with local clumpiness}

In this subsection, 
we fix the value of $\delta$ as 0 and study the effects of local 
clumpiness. 
The vanishing $\delta$ implies that $D(z)$ agrees with the angular
diameter distance of the concordance $\Lambda$CDM model, 
and thus the weak
singularity appears at the symmetry center. 

In the early universe, 
the matter distribution might have been smooth, 
whereas it might have been highly clumpy after
the structures formed. 
The simplest way to take account of the effects of local clumpiness 
and the growth of structures is 
to introduce a $z$-dependent smoothness parameter $\alpha(z)$, which
represents the fraction of matter spreading out almost homogeneously. 
The light rays propagating from standard candles to us pass 
through regions filled
with matter of energy density $\alpha(z)\rho$. 

\begin{figure}[htbp]
\begin{center}
\includegraphics[scale=0.83]{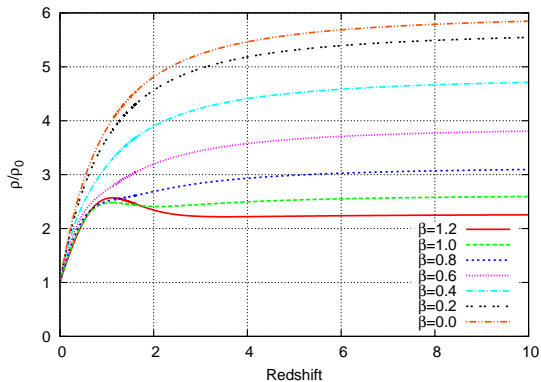}
\caption{Rest-mass densities $\rho(t_0,r(z))$ depicted as 
functions of the cosmological redshift for $\delta=0$ and 
various values of $\beta$. 
}
\label{fig:rho}
\end{center}
\end{figure}

\begin{figure}[htbp]
\begin{center}
\includegraphics[scale=0.83]{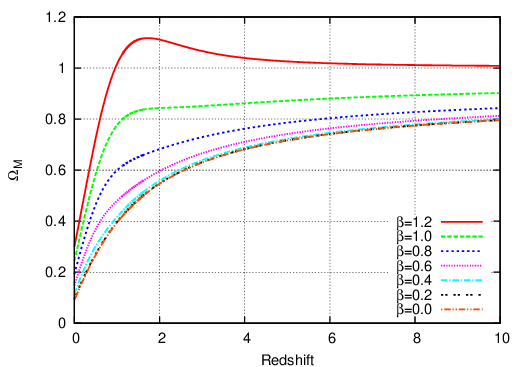}
\caption{Density-parameter functions $\Omega_{\rm M}(t_0,r(z))$ 
depicted as 
functions of the cosmological redshift 
for $\delta=0$ and various values of $\beta$. 
}
\label{fig:Om0}
\end{center}
\end{figure}

\begin{figure}[htbp]
\begin{center}
\includegraphics[scale=0.83]{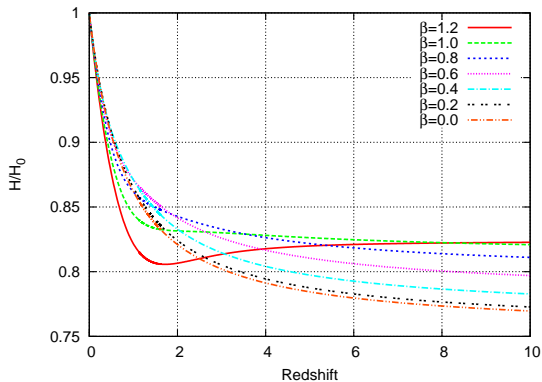}
\caption{Local Hubble function $H(t_0,r(z))$ depicted as 
functions of the cosmological redshift for $\delta=0$ and 
various values of $\beta$. 
}
\label{fig:H0T}
\end{center}
\end{figure}

\begin{figure}[htbp]
\begin{center}
\includegraphics[scale=0.83]{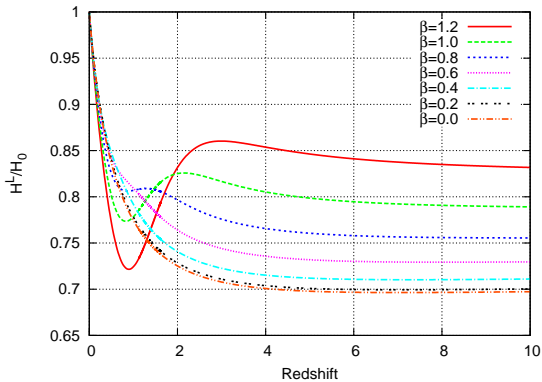}
\caption{Longitudinal expansion rate $H^L(t_0,r(z))$ is depicted as 
functions of the cosmological redshift 
for $\delta=0$ and various values of $\beta$. 
}
\label{fig:H0L}
\end{center}
\end{figure}

At present, the
$z$-dependence of $\alpha(z)$ is still unclear. 
However, as the structures
grow, the spreading-out component of the mass might decrease, 
and therefore $\alpha$ might be an increasing function of $z$. 
Furthermore, there might be a typical redshift $z=\beta$ 
at which almost all of the mass components form clumpy structures. 
This typical redshift $\beta$ will depend on the scenario of 
structure formation. 
For $z<\beta$, the light rays might propagate to us 
through almost empty regions. 
Thus, we assume the following form for the
smoothness parameter $\alpha(z)$:
\begin{equation}
\alpha(z)=1-\exp\left[-z^2/\beta^2\right]. 
\end{equation}
A bundle of light rays propagating through a region where 
$\alpha=1$ is called
a filled beam, 
while a bundle of light rays propagating through a 
region where $\alpha=0$ is called an empty beam. 
Thus, all bundles of light rays
are empty beams in the model of $\beta=\infty$. 
In the case of the empty beam, the angular diameter distance 
does not have a maximal value, and it is clear that 
the distance in the $\beta=\infty$ case cannot fit 
the distance in the $\Lambda$CDM universe 
in all the redshift domain. 
The results depend on the value of $\beta$ as shown in 
Figs. \ref{fig:rho}$-$\ref{fig:H0L}. The resultant 
inhomogeneity is also a large-scale void structure. 
%%%%%%%%%%%%%%%%%%%%%%%%%%%<<start figure>>%%%%%%%%%%%%%%%%%%%%%%%%%%
%%%%%%%%%%%%%%%%%%%%%%%%%%%%<<end figure>>%%%%%%%%%%%%%%%%%%%%%%%%%%%
%%%%%%%%%%%%%%%%%%%%%%%%%%%<<start figure>>%%%%%%%%%%%%%%%%%%%%%%%%%%
%%%%%%%%%%%%%%%%%%%%%%%%%%%%<<end figure>>%%%%%%%%%%%%%%%%%%%%%%%%%%%
%%%%%%%%%%%%%%%%%%%%%%%%%%%<<start figure>>%%%%%%%%%%%%%%%%%%%%%%%%%%
%%%%%%%%%%%%%%%%%%%%%%%%%%%%<<end figure>>%%%%%%%%%%%%%%%%%%%%%%%%%%%
%%%%%%%%%%%%%%%%%%%%%%%%%%%<<start figure>>%%%%%%%%%%%%%%%%%%%%%%%%%%
%%%%%%%%%%%%%%%%%%%%%%%%%%%%<<end figure>>%%%%%%%%%%%%%%%%%%%%%%%%%%%

\begin{figure}[htbp]
\begin{center}
\includegraphics[scale=0.7]{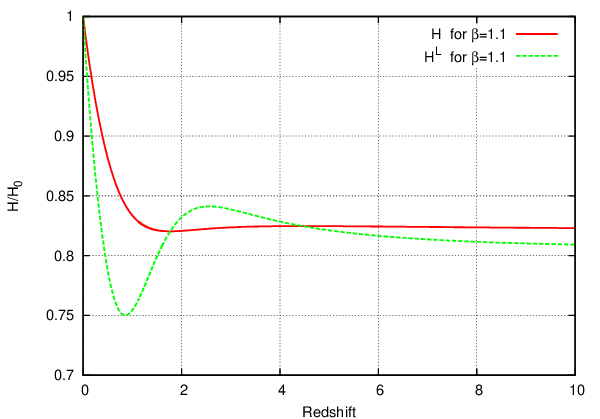}
\caption{Local Hubble function $H(t_0,r(z))$ and 
longitudinal expansion rate $H^L(t_0,r(z))$ depicted as 
functions of the cosmological redshift for $\delta=0$ and $\beta=1.1$. 
}
\label{fig:finetune}
\end{center}
\end{figure}

We show that it is possible to 
reduce the amplitude of the large-scale inhomogeneity by choosing 
an appropriate 
value of $\beta$. In Fig. \ref{fig:finetune}, values of $H(t_0,r(z))$ and 
$H^L(t_0,r(z))$ are depicted as functions of $z$ for $\beta=1.1$. 
In this case, the difference between $H(t_0,r(z))$ and 
$H^L(t_0,r(z))$ is a few percent when $z\gtrsim2$, and  
the amplitude of the inhomogeneity is smaller than 
that depicted in Fig. \ref{fig:0ploth}. 
This means that the appropriate value of the redshift that 
characterizes the period of local clumpiness formation may suppress 
the size of the void by a few Gpc. 
In our model, this appropriate value of the redshift is given by 
$\beta \sim 1.1$. 
In the region $z\gtrsim4$ of our model with $\beta=1.1$, 
the geometry of the universe is almost the same as that of 
the isotropic and homogeneous universe. 
%%%%%%%%%%%%%%%%%%%%%%%%%%%<<start figure>>%%%%%%%%%%%%%%%%%%%%%%%%%%
%%%%%%%%%%%%%%%%%%%%%%%%%%%%<<end figure>>%%%%%%%%%%%%%%%%%%%%%%%%%%%

\subsection{Time variation of the redshift}

\begin{figure}[htbp]
\begin{center}
\includegraphics[scale=1.]{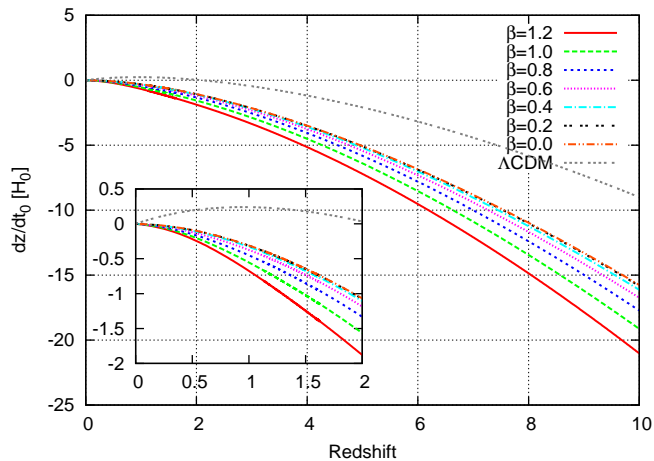}
\caption{Time derivatives of the cosmological redshift for the various models 
depicted as functions of the cosmological redshift. 
}
\label{fig:dzdt}
\end{center}
\end{figure}

We have shown in the preceding subsections that 
it is possible to construct the LTB universe 
model with the same distance-redshift relation as that of the concordance 
$\Lambda$CDM model. 
Thus, it is very important to study how to observationally 
distinguish these two models from each other. 
Here, we show that the 
temporal variation of the cosmological redshift is 
a useful observational 
quantity. 
In Fig. \ref{fig:dzdt}, we depict the derivative $dz/dt_0$ 
of the cosmological redshift with respect to the time $t_0$ of the observer  
at the symmetry center $r=0$ as a function of the cosmological redshift itself. 

As can be seen from this figure, $dz/dt_0$ is 
positive for $0<z\lesssim 2$ in the 
case of the concordance $\Lambda$CDM model, while it is negative for all 
$z$ in the LTB universe models with the uniform big-bang time. 
Therefore, if we observe whether  
$dz/dt_0$ is positive or negative for $z\lesssim2$, 
we can distinguish our LTB model from the concordance $\Lambda$CDM model. 
From Eq. (\ref{eq:dzdt-CDM}), we have
$dz/dt_0|_{z=1}\sim 0.24H_0$ for the concordance $\Lambda$CDM model, and thus, 
the variation of the cosmological redshift in one year is 
$\Delta z|_{z=1} \sim 1.8\times 10^{-11} (H_0$/75 km/s/Mpc). 
Thus, over ten years, 
$\Delta z|_{z=1}$ is larger than 
$10^{-10}$ for the concordance $\Lambda$CDM universe model, and this value 
will become observable in the near future as a result of technological 
innovations\cite{Li:2008cja,2008MNRAS.386.1192L}.

%%%%%%%%%%%%%%%%%%%%%%%%%%%<<start figure>>%%%%%%%%%%%%%%%%%%%%%%%%%%
%%%%%%%%%%%%%%%%%%%%%%%%%%%%<<end figure>>%%%%%%%%%%%%%%%%%%%%%%%%%%%

%%%%%%%%%%%%%%%%%%%%%%%%%%%%%%%%%%%%%%%%%%%%%%%%%%%%%%%%%%%%%%%%
\section{Summary and discussion}
\label{sec6}
%%%%%%%%%%%%%%%%%%%%%%%%%%%%%%%%%%%%%%%%%%%%%%%%%%%%%%%%%%%%%%%%

In this paper, we have attempted 
to solve the inverse problem to construct an LTB 
universe model that has the same distance-redshift relation as that 
of the concordance $\Lambda$CDM model 
with $(\Omega_{\rm M0},\Omega_{\Lambda 0})=(0.3, 0.7)$, 
and we obtained solutions by 
numerical integration. 
In the present study, assuming an inflationary period 
in the early universe, we have restricted ourselves to models with 
uniform big-bang time, and the resultant universe model has a very large 
void whose symmetry center is at the observer's position. 
We have also studied the effects of local clumpiness by 
introducing the smoothness parameter $\alpha$, 
and have shown that the local clumpiness 
may reduce the amplitude of the large-scale inhomogeneity of the void. 
Our results imply that it is possible to construct an 
inhomogeneous but isotropic universe model 
with a distance-redshift relation that agrees quite well 
with the observational data 
of the distance-redshift relation. 

Our LTB universe model is 
regarded as an unnatural model from the viewpoint
of the Copernican principle, 
because the observer stands exactly at the center of the isotropic universe. 
The extent to which we can separate ourselves from 
the center and remain consistent with 
current observations is discussed in 
Refs.\citen{2000ApJ...529...38T} and 
\citen{2000ApJ...529...26T,Tomita:2000rf,Alnes:2006pf,Alnes:2006uk}. 
The strictest limit is given by Alnes and Amarzguioui 
in Ref.\citen{Alnes:2006pf} as 15 Mpc from the observation of CMB. 
This value is much smaller than the cosmological scale, 
and thus we should remain at a special position if the void universe is real. 
However, no observational data has yet been reported that entirely 
excludes inhomogeneous universe models. 
Therefore, it is important to know the type of inhomogeneous universes 
that can explain current observations 
and to propose observational methods for 
testing inhomogeneous universe models. 

In this paper, we have also studied the temporal variation of the 
distance-redshift relation in our LTB universe model whose 
distance-redshift relation is the same as the concordance $\Lambda$CDM model
with $(\Omega_{\rm M0},\Omega_{\Lambda 0})=(0.3, 0.7)$. 
The result implies that if we can observe the time derivative of the 
cosmological redshift with 
sufficient accuracy, we can distinguish our LTB model from 
the concordance $\Lambda$CDM 
universe model. 
Innovations in observational technology might provide us with data on the 
time derivative of the cosmological redshift in the near future
\cite{Uzan:2008qp,Li:2008cja,2008MNRAS.386.1192L}. 

Finally, we should note that if we ignore the inflationary paradigm,  
the functional freedom of the big-bang time $t_{\rm B}(r)$ 
is returned, and we might be able to construct an LTB universe 
model with the same redshift-distance relation and, furthermore, the same 
$dz/dt_0$ as those of the concordance $\Lambda$CDM 
model by choosing an appropriate big-bang time 
$t_{\rm B}(r)$\cite{Uzan:2008qp}. 
However, in this case, we may need an other mechanism to explain the results 
in CMB observations and other 
cosmological problems than the standard cosmology starting from the inflation. 
This will be the subject of a future work and will be discussed elsewhere.

\section*{Acknowledgements}
We are grateful to M. Tanimoto and Y. Nambu for helpful discussions 
and comments. 
This work was supported in part by a JSPS Grant-in-Aid
for Scientific Research (B), No.~17340075. 

\appendix

\section{Characteristics of the Function $\mathcal S(x)$}\label{sec:funcS}

\begin{figure}[htbp]
\begin{center}
\includegraphics[scale=0.83]{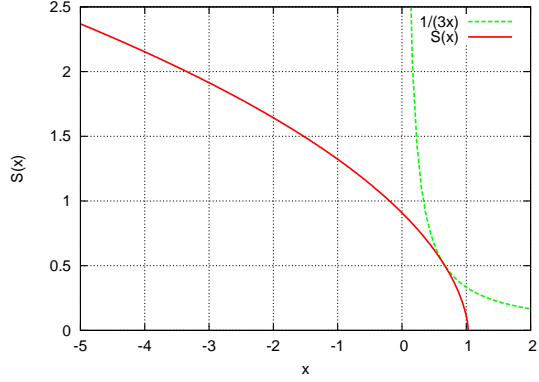}
\caption{$\mathcal S(x)$ and $1/3x$.
}
\label{fig:funcS}
\end{center}
\end{figure}

From Eqs. (\ref{eq:fory}) and (\ref{eq:YS}), 
the function $\mathcal S(x)$ is a solution of 
the following nonlinear differential equation: 
\begin{equation}
4[S(x)+xS'(x)]^2+9x-\frac{3}{S(x)}=0. 
\label{eq:des}
\end{equation}
We can easily see that $S(x)=1/3x$ is also a solution of Eq. (\ref{eq:des}). 
This solution is not equivalent to $S(x)={\mathcal S}(x)$ 
since $\mathcal S(0)=(3/4)^{2/3}$ 
from Eq. (\ref{eq:defS}). 
It is verifiable by Eq. (\ref{eq:defS}) that 
\begin{equation}
x=\tilde x:=\left(\frac{\pi}{6}\right)^{2/3} 
\end{equation}
is a root of the equation 
${\mathcal S}(x)=1/3x$. In other words, the solutions 
$S(x)=1/3x$ and $S(x)={\mathcal S}(x)$ agree with each other at 
$x=\tilde x$ (See Fig. \ref{fig:funcS}).
%%%%%%%%%%%%%%%%%%%%%%%%%%%<<start figure>>%%%%%%%%%%%%%%%%%%%%%%%%%%
%%%%%%%%%%%%%%%%%%%%%%%%%%%%<<end figure>>%%%%%%%%%%%%%%%%%%%%%%%%%%%

Let us consider the behaviour of a solution $S(x)$ that is regular in the 
neighborhood of $x=0$. Expanding $S(x)$ around $x=0$, we have
\begin{equation}
S(x)=a_0+a_1x+\mathcal O(x^2). 
\end{equation}
Substituting this expression into 
Eq. (\ref{eq:des}), we have, from the zeroth order of $x$,   
\begin{equation}
a_0=\left(\frac{3}{4}\right)^{1/3}
\end{equation}
and, from the first order of $x$, 
\begin{equation}
a_1=-\frac{9}{10}6^{-1/3}.
\end{equation}
Therefore, a solution $S(x)$ that is regular at $x=0$ should satisfy 
$S(0)=(3/4)^{2/3}$. 
This result also means that $\mathcal S(x)$ is a unique 
solution that is regular at $x=0$. 

Next, we consider a solution $S(x)$ 
that satisfies $S({\tilde x})=1/3{\tilde x}$.
Expanding $S(x)$ around $x=\tilde x$, we have
\begin{equation}
S=b_0+b_1(x-\tilde x)+b_2(x-\tilde x)^2+\mathcal O((x-\tilde x)^3). 
\end{equation}
Substituting this expression into Eq. (\ref{eq:des}), we find, 
from the zeroth order of $(x-\tilde x)$, that
\begin{equation}
b_1=-\frac{1}{3\tilde x^2}.
\end{equation}
Then, we have 
\begin{equation}
\left.S'(x)\right|_{x=\tilde x}=b_1=-\frac{1}{3\tilde x^2}
=\left.\left(\frac{1}{3x}\right)'\right|_{x=\tilde x}. 
\end{equation}
Therefore, the derivative of $S(x)$ at $x=\tilde x$ is unique. This result implies 
that the solutions $S(x)=1/3x$ and $S(x)=\mathcal S(x)$ have the same gradient 
at $x=\tilde x$.
The first order of $(x-\tilde x)$ in Eq. (\ref{eq:des}) 
is automatically satisfied. 
From the second order, we have 
\begin{equation}
\left(b_2-\frac{1}{3\tilde x^3}\right)
\left(b_2-\frac{1}{3\tilde x^3}+\frac{27}{16}\right)=0. 
\end{equation}
There are two roots $b_2=b_{2\pm}$ of the above equation, where 
\begin{equation}
b_{2+}=\frac{1}{3{\tilde x}^3} ~~~~~~{\rm and}~~~~~~b_{2-}=\frac{1}{3{\tilde x}^3}-\frac{27}{16}.
\end{equation}
This fact means that there are at least 
two solutions of Eq. (\ref{eq:des}) with 
the same value and the same derivative at $x=\tilde x$. 
The root $b_2=b_{2+}$ corresponds to the solution $S(x)=1/3x$ whereas 
the other root $b_2=b_{2-}$ corresponds to the solution $S(x)=\mathcal S(x)$. 
Since we have 
\begin{equation}
\mathcal S+x\mathcal S'=2(b_1+\tilde x b_{2-})(x-\tilde x)
+\mathcal O((x-\tilde x)^2), 
\end{equation}
and since $b_1<0$ and $b_{2-}<0$, we have 
\begin{eqnarray}
\mathcal S+x\mathcal S'&>&0~~{\rm for}~~x<\tilde x, \\
\mathcal S+x\mathcal S'&<&0~~{\rm for}~~x>\tilde x. 
\end{eqnarray}
Eventually, we obtain the differential equation
for $\mathcal S(x)$ as
\begin{eqnarray}
\mathcal S+x\mathcal S'&=&
\frac{\sqrt{3}}{2}\sqrt{\frac{1}{\mathcal S}-3x} 
~~ {\rm for} ~~x<\tilde x, \\
\mathcal S+x\mathcal S'&=&
-\frac{\sqrt{3}}{2}\sqrt{\frac{1}{\mathcal S}-3x}
~~ {\rm for}~~ x>\tilde x.
\end{eqnarray}

\section{Dyer-Roeder Equation}
\label{sec:DRdis}

Let us consider a bundle of light rays 
whose sectional area is given by $A$. 
The expansion rate $\theta$ of $A$ along the ray is defined by 
\begin{equation}
\theta=\frac{\dot A}{2A}, \label{eq:deftheta}
\end{equation}
where the dot denotes differentiation with respect to an 
affine parameter $\lambda$ of the ray bundle. 
Then the evolution of $\theta$ is given by \cite{1961RSPSA.264..309S}
\begin{equation}
\dot \theta=-\theta^2-\sigma^2-\frac{1}{2}R_{\mu\nu}p^\mu p^\nu, 
\label{eq:theta-eq}
\end{equation}
where $p^\mu$ is a tangent vector  
of the light ray, and $\sigma^2$ is the shear factor defined by 
\begin{equation}
\sigma^2=\frac{1}{2}\nabla_\mu p_\nu \nabla^\mu p^\nu
-\frac{1}{4}(\nabla_\mu p^\mu)^2. 
\end{equation}
Substituting Eq. (\ref{eq:deftheta}) into Eq. (\ref{eq:theta-eq}), 
we have
\begin{equation}
\ddot{\sqrt{A}}=-\left(\sigma^2+\frac{1}{2}R_{\mu\nu}p^\mu p^\nu\right)
\sqrt{A}. 
\label{eq:difarea}
\end{equation}

Let us assume that
the light beam remains far away from clumps and that 
the contribution of $\sigma$ in Eq. (\ref{eq:difarea}) is negligible 
\cite{1969ApJ...155...89K,1973ApJ...180L..31D,Kantowski1998}. 
In addition, we assume that the fraction of matter density 
that is smoothly distributed is given by 
$\alpha$, namely, the energy density on the ray is given by 
$\alpha\rho$. 
As mentioned in \S1, $\alpha$ is the so-called smoothness parameter. 
Then, Eq. (\ref{eq:difarea}) reduces to 
\begin{equation}
\ddot{\sqrt{A}}=-4\pi\dot t^2\alpha \rho\sqrt{A}, 
\label{eq:difarea2}
\end{equation}
where we have used the Einstein equation with 
the energy momentum tensor (\ref{eq:enmomedust}). 

Since the relation between the angular diameter distance 
and $A$ is given by
\begin{equation}
D\propto\sqrt{A}, 
\end{equation}
we obtain 
\begin{equation}
\ddot D=-4\pi\dot t^2\alpha \rho D. 
\end{equation}
When $D$ is given as a function of redshift $z$, 
we can rewrite the above equation in the form of Eq. (\ref{eq:dreq}). 

\section{Derivation of Basic Equations}\label{sec:cal}
From Eq.
(\ref{eq:defz}), we have
\begin{equation}
\dot r=\frac{1+z}{H_0}. 
\end{equation}
This is identical to Eq. (\ref{eq:dchi}). 
From Eq. (\ref{eq:defzeta}), 
we obtain 
\begin{equation}
\dot z=\frac{\zeta}{dD/dz} 
\end{equation}
as Eq. (\ref{eq:dz}). 
Using $\zeta$, 
Eq. (\ref{eq:dreq}) can be written as Eq. (\ref{eq:dzeta}):
\begin{equation}
\dot \zeta=-4\pi\frac{(1+z)^2}{H_0^2}\alpha\rho D. 
\label{eq:dreq2}
\end{equation}

The remaining task is the derivation of Eqs. (\ref{eq:dm}) 
and (\ref{eq:dk})
from Eqs. (\ref{eq:nullcon}) and (\ref{eq:nullgeo}). 
Differentiating Eq. (\ref{eq:YS}) with respect to $r$, we have
\begin{eqnarray}
\partial_r R
&=&\frac{1}{3}rm^{-2/3}(t-t_{\rm B})^{2/3}
\left(\mathcal S-2x\mathcal S'\right)\frac{\dot m}{\dot r}\nonumber\\
&-&
\frac{2}{3}rm^{1/3}(t-t_{\rm B})^{-1/3}
\left(\mathcal S+x\mathcal S'\right)\frac{\dot t_{\rm B}}{\dot r}\nonumber\\
&+&
rm^{-1/3}(t-t_{\rm B})^{4/3}\mathcal S'\frac{\dot k}{\dot r}\nonumber\\
&+&
m^{1/3}(t-t_{\rm B})^{2/3}\mathcal S. 
\label{eq:dcY}
\end{eqnarray}
Substituting Eqs. (\ref{eq:dcY}) and (\ref{eq:defz}) 
into Eq. (\ref{eq:nullcon})Cwe have
\begin{equation}
a\dot m+b\dot t_{\rm B}+c\dot k+d=0, 
\label{eq:abcd}
\end{equation}
where
\begin{eqnarray}
a&=&\frac{1}{3}rm^{-2/3}(t-t_{\rm B})^{2/3}
\left(\mathcal S-2x\mathcal S'\right), \\
b&=&-\frac{2}{3}rm^{1/3}(t-t_{\rm B})^{-1/3}
\left(\mathcal S+x\mathcal S'\right), \\
c&=&rm^{-1/3}(t-t_{\rm B})^{4/3}\mathcal S', \\
d&=&-\frac{1+z}{H_0}\left[\sqrt{1-kr^2}
-m^{1/3}(t-t_{\rm B})^{2/3}\mathcal S\right]. 
\end{eqnarray}

Using Eqs. (\ref{eq:gaucon}) and (\ref{eq:defz}), 
we can rewrite Eq. (\ref{eq:nullgeo}) as 
\begin{equation}
-\partial_r R \frac{\zeta}{H_0 dD/dz}
+\partial_t\partial_r R \left(\frac{1+z}{H_0}\right)^2=0. 
\label{eq:nullgeo2}
\end{equation}
Differentiating Eq. (\ref{eq:dcY}) with respect to $t$, 
we have
\begin{eqnarray}
\partial_t\partial_r R
&=&\frac{1}{6}rm^{-2/3}(t-t_{\rm B})^{-1/3}
\frac{1}{\mathcal S^2}
\frac{\dot m}{\dot r}\nonumber\\
&+&\frac{1}{6}rm^{1/3}(t-t_{\rm B})^{-4/3}
\frac{1}{\mathcal S^2}
\frac{\dot t_{\rm B}}{\dot r}\nonumber\\
&+&\frac{2}{3}rm^{-1/3}(t-t_{\rm B})^{1/3}
\left(2\mathcal S'+x\mathcal S''\right)\frac{\dot k}{\dot r}\nonumber\\
&+&\frac{2}{3}m^{1/3}(t-t_{\rm B})^{-1/3}(\mathcal S+x\mathcal S'), 
\label{eq:dtcY}
\end{eqnarray}
where we have used the following equation for $\mathcal S(x)$
(see Ref.\citen{Tanimoto:2007dq}):
\begin{equation}
x(2\mathcal S\mathcal S''+\mathcal S'^2)+5\mathcal S\mathcal S'+\frac{9}{4}=0. 
\end{equation}

Substituting Eqs. (\ref{eq:nullcon}) and (\ref{eq:dtcY}) 
into Eq. (\ref{eq:nullgeo2}), we have 
\begin{equation}
e\dot m+f\dot t_{\rm B}+g\dot k+h=0, 
\label{eq:efgh}
\end{equation}
where
\begin{eqnarray}
e&=&rm^{-2/3}(t-t_{\rm B})^{-1/3}, \\
f&=&rm^{1/3}(t-t_{\rm B})^{-4/3}, \\
g&=&4rm^{-1/3}(t-t_{\rm B})^{1/3}\mathcal S^2
\left(2\mathcal S'+x\mathcal S''\right), \\
h&=&-\frac{6\sqrt{1-kr^2}\zeta}{(1+z)dD/dz}\mathcal S^2
+\frac{1+z}{H_0}
\left[4m^{1/3}(t-t_{\rm B})^{-1/3}
\mathcal S^2(\mathcal S+x\mathcal S')\right]. 
\end{eqnarray}

If $t_{\rm B}(r)$ is given, from Eqs. (\ref{eq:abcd}) and (\ref{eq:efgh}), 
we have
\begin{eqnarray}
\dot m&=&F_m(m,k,r,z,\zeta):=\frac{c\tilde h-\tilde dg}{ag-ce}, \\
\dot k&=&F_k(m,k,r,z,\zeta):=\frac{\tilde de-a\tilde h}{ag-ce}, 
\end{eqnarray}
where
\begin{eqnarray}
\tilde d&=&d+\frac{b(1+z)\partial_r t_{\rm B}}{H_0}, \\
\tilde h&=&h+\frac{f(1+z)\partial_r t_{\rm B}}{H_0}. 
\end{eqnarray}

\section{\bm $(\partial_r^2R)_{\rm b}$, 
$(\partial_t^2\partial_rR)_{\rm b}$ and 
$(\partial_t\partial_r^2R)_{\rm b }$}
\label{sec:delrs}

We show how to calculate $\partial_r^2R$, $\partial_t^2\partial_rR$ and 
$\partial_t\partial_r^2R$ on the null geodesic that 
corresponds to the observed light ray. 
Let us consider the differentiation 
of the null condition (\ref{eq:nullcon}) along 
the null geodesic $(t,r)=(t_{\rm b},r_{\rm b})$ 
with respect to $\lambda$ as follows:
\begin{equation}
\frac{d}{d\lambda}\left(\partial_r R-\sqrt{1-kr^2}\right)=0. 
\end{equation}
Using $\dot t=-\dot r$, we have
\begin{equation}
(\partial_r^2R)_{\rm b}
=(\partial_t\partial_rR)_{\rm b}
-\left(\frac{\partial_r(kr^2)}{2\sqrt{1-kr^2}}\right)_{\rm b}, 
\end{equation}
where $\partial_t\partial_rR$ is given by Eq. (\ref{eq:dtcY}). 

The expression for $\partial_t^2\partial_rR$ 
is given by differentiating Eq. (\ref{eq:fory}) 
with respect to $t$ and $r$ as 
\begin{equation}
\partial_t^2\partial_rR=-\frac{\partial_r M}{R^2}
+\frac{2M\partial_rR}{R^3},  
\label{eq:D2}
\end{equation}
where $\partial_rR$ is given by Eq. (\ref{eq:dcY}).

Finally, $(\partial_t\partial_r^2R)_{\rm b}$ is given by 
differentiating the null geodesic equation (\ref{eq:nullgeo}) 
with respect to the affine parameter $\lambda$. 
Namely, 
\begin{eqnarray}
&&\frac{d}{d\lambda}
\left(\partial_rR\ddot t+\partial_t\partial_rR\dot t^2\right)=0.
\end{eqnarray}
The above equation leads to
\begin{equation}
(\partial_t\partial_r^2R)_{\rm b}
=(\partial_t^2\partial_rR)_{\rm b}
+\dot t^{-3}_{\rm b}\left(\partial_rR\dddot t
+3\partial_t\partial_rR\dot t\ddot t
-\partial_r^2R\dot t\ddot t\right)_{\rm b}, 
\end{equation}
where 
\begin{eqnarray}
\dot t&=&-\frac{1+z}{H_0}, \\
\ddot t&=&-\frac{\zeta}{H_0dD/dz}, \\
\dddot t&=&\frac{4\pi(1+z)^2\alpha\rho D}{H_0^3dD/dz}
+\frac{\zeta^2 d^2D/dz^2}{H_0(dD/dz)^3}. 
\end{eqnarray}

\section{Initial Conditions at the Center without Distance Modification}
\label{sec:iniv}

From the null condition, we have 
$\partial_r R\sim 1$ near the center. 
Thus we must have $R\sim r$ in order that the metric (\ref{eq:metric})
is regular at the center. 
Therefore, from Eq. (\ref{eq:rho}), 
$M$ should be given by 
\begin{equation}
M(r)=\frac{m_0}{6}r^3+\mathcal O(r^4)
\end{equation}
for the finiteness of $\rho$ at the center. 
Thus, we have 
\begin{equation}
m(r)=m_0+\mathcal O(r). 
\end{equation}
Hereafter, the subscript $0$ describes the initial value at the center.
Obviously, $r_0=z_0=0$. 
The remaining initial values are $m_0$, $k_0$ and $\zeta_0$. 
In addition, the initial value of $t_0$ should also be determined. 

Since 
\begin{equation}
R\sim r
\end{equation}
near the center, Eqs. (\ref{eq:fory}), (\ref{eq:rho}) and (\ref{eq:YS}) are 
given by
\begin{equation}
8\pi\rho_0=m_0,
\label{eq:rhom}
\end{equation}
\begin{equation}
\left.\left(\frac{\partial_t R}{R}\right)^2\right|_{r=0,t=t_0}
=H_0^2=-k_0+\frac{8\pi\rho_0}{3}, 
\label{eq:okom}
\end{equation}
\begin{equation}
x_0\mathcal S(x_0)=\frac{k_0}{8\pi\rho_0}, 
\label{eq:y0}
\end{equation}
where
\begin{equation}
x_0=k_0\left(\frac{t_0}{m_0}\right)^{2/3}. 
\end{equation}
Equation (\ref{eq:okom}) can be rewritten as 
\begin{equation}
1=\Omega_{\rm k0}+\Omega_{\rm M0}, 
\label{eq:omegas}
\end{equation}
where 
\begin{eqnarray}
\Omega_{\rm k0}&=&-k_0/H_0^2, \\ 
\Omega_{\rm M0}&=&8\pi\rho_0/(3H_0^2). \label{eq:Om0}
\end{eqnarray}

Once we fix the value of $\Omega_{\rm m0}$ at the symmetry center, 
$\Omega_{\rm k0}$, or equivalently $k_0$, 
is fixed by Eq. (\ref{eq:omegas}). 
$m_0$ and $\rho_0$ are related to $\Omega_{\rm m0}$ 
by Eqs. (\ref{eq:rhom}) and (\ref{eq:Om0}). 
$t_0$ is given as the solution of Eq. (\ref{eq:y0}). 
The remaining initial value is $\zeta_0=\dot z_0 dD/dz|_{z=0}=\dot z_0/H_0$. 
Differentiating Eq. (\ref{eq:fory}) with respect to $r$, 
we have 
\begin{equation}
\partial_t\partial_r R=\frac{1}{2\partial_t R}\left(-(\partial_r k) r^2
-2kr+\frac{2\partial_r M}{R}-\frac{2M\partial_r R}{R^2}\right). 
\end{equation}
The above equation becomes 
\begin{equation}
\partial_t\partial_r R=H_0
\end{equation}
at $r=0$ and $t=t_0$. 
Using this equation, 
from Eqs. (\ref{eq:defz}) and (\ref{eq:nullgeo}), 
we obtain
\begin{equation}
\dot z_0=
\frac{1}{H_0}\left.\frac{\partial_t\partial_r R}{\partial_r R}
\right|_{r=0,t=t_0}=1. 
\end{equation}
Thus, we have
\begin{equation}
\zeta_0=\dot z_0 \left.\frac{dD}{dz}\right|_{z=0}
=\frac{\dot z_0}{H_0}=\frac{1}{H_0}. 
\end{equation}

%\bibliographystyle{prsty}
%\bibliography{hybriden}

\end{document}